\documentclass[iop,apj]{emulateapj}
\usepackage{apjfonts}
\usepackage{lineno}
\usepackage[normalem]{ulem}
\usepackage{verbatim}
\usepackage{ulem}
\usepackage{graphicx}
\usepackage{hyperref}
\usepackage{ifpdf}

\shortauthors{CROMARTIE ET AL.}
\shorttitle{NEW ARECIBO MILLISECOND PULSARS IN \emph{FERMI} SOURCES}

\begin{document}

\def\fermi{{\em Fermi}}
\def\swift{{\em Swift}}
\def\cxo{{\em CXO}}
\def\hr{^{\rm h}}
\def\mi{^{\rm m}}
\def\s{^{\rm s}}
\def\msun{M$_{\sun}$}

\def\psra{PSR~J0251$+$26}
\def\psrb{PSR~J1048$+$2339}
\def\psrc{PSR~J1805$+$06}
\def\psrd{PSR~J1824$+$10}
\def\psre{PSR~J1909$+$21}
\def\psrf{PSR~J2052$+$1218}
\def\psrg{PSR~J1921$+$01}
\def\psrh{PSR~J2042$+$02}

\title{Six New Millisecond Pulsars from Arecibo Searches of {\em Fermi \/} Gamma-Ray Sources}

\author{
H.~T.~Cromartie\altaffilmark{1,2}, 
F.~Camilo\altaffilmark{3}, 
M.~Kerr\altaffilmark{4}, 
J.~S.~Deneva\altaffilmark{5},
S.~M.~Ransom\altaffilmark{6},
P.~S.~Ray\altaffilmark{5},
E.~C.~Ferrara\altaffilmark{7},
P.~F.~Michelson\altaffilmark{8}, 
and K.~S.~Wood\altaffilmark{5}}

\altaffiltext{1}{Department of Astronomy, University of Virginia, Charlottesville, VA 22903, USA} 
\altaffiltext{2}{email: thankful@virginia.edu}
\altaffiltext{3}{Columbia Astrophysics Laboratory, Columbia University, New York, NY 10027, USA}
\altaffiltext{4}{CSIRO Astronomy and Space Science, Australia Telescope National Facility, Epping, NSW~1710, Australia}
\altaffiltext{5}{Space Science Division, Naval Research Laboratory, Washington, DC 20375-5352, USA}
\altaffiltext{6}{National Radio Astronomy Observatory (NRAO), Charlottesville, VA 22903, USA}
\altaffiltext{7}{NASA Goddard Space Flight Center, Greenbelt, MD 20771, USA}
\altaffiltext{8}{W. W. Hansen Experimental Physics Laboratory, Kavli Institute for Particle Astrophysics and Cosmology, Department of Physics and SLAC National Accelerator Laboratory, Stanford University, Stanford, CA 94305, USA}

\begin{abstract}
We have discovered six radio millisecond pulsars (MSPs) in a search with the Arecibo telescope of 34 unidentified gamma-ray sources from the \fermi{} Large Area Telescope (LAT) 4-year point source catalog. Among the 34 sources, we also detected two MSPs previously discovered elsewhere. Each source was observed at a center frequency of 327 MHz, typically at three epochs with individual integration times of 15 minutes. The new MSP spin periods range from 1.99 to 4.66 ms. Five of the six pulsars are in interacting compact binaries (period $\leq$ 8.1 hr), while the sixth is a more typical neutron star-white dwarf binary with an 83-day orbital period. This is a higher proportion of interacting binaries than for equivalent \fermi{}-LAT searches elsewhere. The reason is that Arecibo's large gain afforded us the opportunity to limit integration times to 15 minutes, which significantly increased our sensitivity to these highly accelerated systems. Seventeen of the remaining 26 gamma-ray sources are still categorized as strong MSP candidates, and will be re-searched.
\end{abstract}


\section{Introduction} \label{sec:intro}
Of the 230 millisecond pulsars (MSPs) currently known in the Galactic disk\footnote{Public list of Galactic MSPs: \url{http://astro.phys.wvu.edu/GalacticMSPs/}}, 30\% have been discovered in previously unidentified sources of gamma rays detected by the \fermi{}-LAT instrument\footnote{For a list of all LAT pulsars, see \url{https://confluence.slac.stanford.edu/display/GLAMCOG/Public+List+of+LAT-Detected+Gamma-Ray+Pulsars}} \citep[][]{aaa+09m}. While only around 10\% of all known pulsars rotate at millisecond rates, MSPs make up half of all pulsars observed to emit gamma rays \citep{car14}. The LAT source catalogs have been instrumental in the search for new MSPs, providing spectral data to aid in distinguishing possible MSPs from other gamma-ray-emitting objects, such as active galactic nuclei (AGNs). Once an MSP is discovered in a radio search and a phase-connected timing solution is available, the sparse gamma-ray photons are folded using the radio ephemeris in order to glimpse gamma-ray pulsations \citep[e.g.,][]{cgj+11}. While it was possible to search for radio pulsars in gamma-ray sources prior to the \fermi{} era \citep[e.g.,][]{cml05,rob02}, the small positional uncertainty of LAT gamma-ray sources has enabled single-pointing radio searches. Overall, the search for MSPs in the Galactic disk has been made extremely efficient by employing \fermi-LAT data in selecting radio search targets.

Before 2013, no \fermi{} MSPs had been discovered using the 305-m Arecibo radio telescope in Puerto Rico. In contrast, the Green Bank (GBT), Parkes, Nan\c{c}ay, Giant Metrewave (GMRT), and Effelsberg telescopes had been used to discover dozens of new MSPs using the 1FGL and 2FGL catalogs \citep{aaa+10g,naa+12} as guides. In this work, we present the first six MSPs discovered in unidentified LAT sources using the Arecibo telescope, along with preliminary orbital parameters gleaned from radio timing. We then quantitatively discuss the relative sensitivities --- both in the flux density and acceleration regimes --- between the \fermi{}-LAT MSP searches conducted at Arecibo and those done at the GBT and Parkes.

\section{Candidate Selection} \label{sec:cands}
All but two of the 34 candidates observed came directly from early versions of the third \fermi-LAT catalog (3FGL, also known as the 4-year catalog), which was later published by \cite{aaa+15}. The two remaining sources were detected with the LAT but were below the significance threshold required for inclusion in the final catalog. While LAT source lists contain $>$1000 unidentified gamma-ray emitters, several constraints dramatically limit the number of sources appropriate for our searches.

Every source had to be located within the declination range of Arecibo ($-1^{\circ} < \delta < 39^{\circ}$). Justification for picking the 327-MHz receiver over L-band, for example, was two-fold. First, the target source error circles were required to fit within Arecibo's beam, allowing for single, long-duration pointings. The 327-MHz system, with a relatively large $\mbox{FWHM}=15'$, was the best choice. Second, pulsars have steep spectra, and therefore are brighter at such a relatively low frequency. Very few pulsars are known to be variable in gamma rays \citep{rap+12}; thus, only non-variable LAT sources were selected. Also, each of the selected sources had a spectral energy distribution consistent with those of known gamma-ray pulsars, which typically have exponentially cut-off power-law spectra \citep{aaa+13}. Because it is difficult to characterize LAT sources amid the Galaxy's diffuse gamma-ray background \citep[see, e.g.,][]{gk2012}, and because the effects of dispersion, scattering, and synchrotron emission inhibit radio pulsar observations at low frequency along the Galactic plane, only sources with $|b|>5^\circ$ were considered. After whittling the list down, we obtained the 34 sources in Table~\ref{tab:survey}.

\begin{deluxetable*}{lccrrcr}
\tablewidth{0pt}
\tablecaption{\label{tab:survey} Summary of Arecibo Searches of Unidentified \fermi-LAT Sources}
\tablecolumns{7}
\tablehead{
\colhead{Name}                  &
\colhead{R.A.\tablenotemark{a}}                  &
\colhead{Decl.\tablenotemark{a}}                 &
\colhead{$l$}                                    &
\colhead{$b$}                                    &
\colhead{Integration Time}                       &
\colhead{$\mbox{DM}_{\rm max}$\tablenotemark{b}} \\
\colhead{}                &
\colhead{(J2000.0)}       &
\colhead{(J2000.0)}       &
\colhead{(deg)}           &
\colhead{(deg)}           &
\colhead{(minutes)}       &
\colhead{(pc\,cm$^{-3}$)}
}
\startdata
3FGL~J0103.7$+$1323 & $01\hr03\mi46\s$ & $13\arcdeg23'33\arcsec$ & 127.5 & $-$49.4 & 15, 15, 15 & 72\\
3FGL~J0134.5$+$2638 & $01\hr34\mi31\s$ & $26\arcdeg38'15\arcsec$ & 134.7 & $-$35.2 & 15, 15, 15 & 92\\
3FGL~J0232.9$+$2606 & $02\hr32\mi56\s$ & $26\arcdeg06'13\arcsec$ & 149.7 & $-$31.4 & 15, 15, 15 & 130\\
\textbf{3FGL~J0251.1$+$2603} & $02\hr51\mi08\s$ & $26\arcdeg04'48\arcsec$ & 153.9 & $-$29.5 & \textbf{15, 15} & 124\\
3FGL~J0318.1$+$0252 & $03\hr18\mi09\s$ & $02\arcdeg52'10\arcsec$ & 178.4 & $-$43.6 & 15, 15, 15 & 900\\
3FGL~J0330.6$+$0437 & $03\hr30\mi40\s$ & $04\arcdeg37'32\arcsec$ & 179.5 & $-$40.1 & 15, 15, 15 & 680\\
3FGL~J0342.3$+$3148c & $03\hr42\mi18\s$ & $31\arcdeg48'33\arcsec$ & 160.3 & $-$18.4 & 15, 15, 15 & 100\\
3FGL~J0421.6$+$1950 & $04\hr21\mi37\s$ & $19\arcdeg50'49\arcsec$ & 175.9 & $-$20.7 & 15, 15, 15 & 150\\
3FGL~J0517.1$+$2628c & $05\hr17\mi10\s$ & $26\arcdeg28'44\arcsec$ & 178.6 & $-$6.6 & 10, 15, 15, 15, 15 & 120\\
3FGL~J0539.8$+$1434 & $05\hr39\mi48\s$ & $14\arcdeg33'53\arcsec$ & 191.6 & $-$8.6 & 5, 5 & 200\\
\textbf{3FGL~J1048.6$+$2338} & $10\hr48\mi41\s$ & $23\arcdeg38'29\arcsec$ & 213.2 & 62.1 & 15, \textbf{15} & 66\\
3FGL~J1049.7$+$1548 & $10\hr49\mi44\s$ & $15\arcdeg48'25\arcsec$ & 228.5 & 59.6 & 10, 15 & 68\\
3FGL~J1200.4$+$0202 & $12\hr00\mi27\s$ & $02\arcdeg02'31\arcsec$ & 274.8 & 62.1 & 15, 15, 15 & 64\\
3FGL~J1225.9$+$2953 & $12\hr25\mi59\s$ & $29\arcdeg53'25\arcsec$ & 185.2 & 83.8 & 15, 15, 15 & 40\\
P7R4~J1250$+$3118\tablenotemark{e} & $12\hr50\mi52\s$ & $31\arcdeg18'18\arcsec$ & 124.6 & 85.8 & 15, 15, 15 & 40\\
3FGL~J1309.0$+$0347 & $13\hr09\mi02\s$ & $03\arcdeg47'27\arcsec$ & 313.9 & 66.3 & 15, 15, 10, 15, 10, 15 & 60\\
3FGL~J1322.3$+$0839 & $13\hr22\mi20\s$ & $08\arcdeg39'27\arcsec$ & 325.9 & 70.1 & 15, 15, 15 & 52\\
3FGL~J1601.9$+$2306 & $16\hr01\mi57\s$ & $23\arcdeg06'39\arcsec$ & 38.5 & 46.9 & 15, 15, 15 & 60\\
3FGL~J1627.8$+$3217 & $16\hr27\mi52\s$ & $32\arcdeg17'56\arcsec$ & 53.0 & 43.2 & 15, 15 & 70\\
3FGL~J1704.1$+$1234 & $17\hr04\mi08\s$ & $12\arcdeg34'25\arcsec$ & 32.5 & 29.4 & 15, 15, 15 & 116\\
3FGL~J1720.7$+$0711 & $17\hr20\mi46\s$ & $07\arcdeg11'21\arcsec$ & 29.0 & 23.4 & 15, 15, 15 & 156\\
\textbf{3FGL~J1805.9$+$0614} & $18\hr05\mi55\s$ & $06\arcdeg14'15\arcsec$ & 33.4 & 13.0 & \textbf{15, 15} & 304\\
\textbf{3FGL~J1824.0$+$1017} & $18\hr24\mi05\s$ & $10\arcdeg17'27\arcsec$ & 39.1 & 10.7 & \textbf{15, 15} & 356\\
3FGL~J1827.7$+$1141 & $18\hr27\mi42\s$ & $11\arcdeg41'50\arcsec$ & 40.8 & 10.5 & 15, 15 & 356\\
3FGL~J1829.2$+$3229 & $18\hr29\mi08\s$ & $32\arcdeg30'42\arcsec$ & 60.7 & 18.5 & 15, 15, 15 & 158\\
3FGL~J1842.2$+$2742 & $18\hr42\mi15\s$ & $27\arcdeg42'09\arcsec$ & 57.1 & 14.1 & 15, 15, 15 & 216\\
\textbf{P7R4~J1909$+$2102\tablenotemark{e}} & $19\hr09\mi32\s$ & $21\arcdeg02'56\arcsec$ & 53.7 & 5.6 & 15, \textbf{15} & 564\\
\textbf{3FGL~J1921.2$+$0136\tablenotemark{d}} & $19\hr21\mi14\s$ & $01\arcdeg36'26\arcsec$ & 37.8 & $-$5.9 & \textbf{5, 10, 15, 15} & 670\\
3FGL~J2026.3$+$1430 & $20\hr26\mi21\s$ & $14\arcdeg30'53\arcsec$ & 57.3 & $-$13.4 & 15, 15, 15 & 226\\
\textbf{3FGL~J2042.1$+$0247\tablenotemark{d}}  & $20\hr42\mi09\s$ & $02\arcdeg47'35\arcsec$ & 49.0 & $-$23.0 & \textbf{15, 15} & 140\\
\textbf{3FGL~J2052.7$+$1217} & $20\hr52\mi47\s$ & $12\arcdeg17'51\arcsec$ & 59.1 & $-$20.0 & \textbf{15, 15, 15} & 148\\
3FGL~J2108.0$+$3654 & $21\hr08\mi02\s$ & $36\arcdeg55'19\arcsec$ & 81.1 & $-$7.2 & 15, 15, 15 & 360\\
3FGL~J2212.5$+$0703 & $22\hr12\mi35\s$ & $07\arcdeg03'35\arcsec$ & 68.7 & $-$38.6 & 15, 30, 15, 15, 15, 5 & 84\\
3FGL~J2352.0$+$1752 & $23\hr52\mi04\s$ & $17\arcdeg52'50\arcsec$ & 103.5 & $-$42.7 & 15, 15, 15, 15 & 74\\
\enddata
\tablecomments{Boldfaced entries denote observations yielding MSP detections.}
\tablenotetext{a}{Arecibo telescope pointing position.}
\tablenotetext{b}{The maximum dispersion measure (DM) up to which we searched corresponds approximately to twice the maximum value predicted for each line of sight by the NE2001 electron density model \citep{cl02}, with the exception of 3FGL J0318.1+0252 and J0330.6+0437, which were unintentionally searched to higher DMs.}
\tablenotetext{d}{Discovered at the GBT (S.\ Sanpa-Arsa et al.\ 2016, in preparation).} 
\tablenotetext{e}{Source not included in 3FGL catalog.}
\end{deluxetable*}

\section{Observations and Data Analysis} \label{sec:obs}
Observations of the 34 \fermi{}-LAT sources were conducted during 12 sessions between 2013 June and September using the Arecibo telescope. In order to combat the effects of scintillation, orbital acceleration, and eclipses (discussed further below), we aimed to observe each source three times for 15 minutes per pointing, though the exact number of pointings per source changed as data were analyzed.

Sessions in the months of June, July and early August were conducted in-person at the observatory, while later sessions occurred remotely. In either case, the standard CIMA\footnote{\url{http://www.naic.edu/~cima}} telescope control software was used in conjunction with command-line control of the Puertorican Ultimate Pulsar Processing Instrument (PUPPI). The PUPPI backend (a replica of GUPPI\footnote{\url{https://safe.nrao.edu/wiki/bin/view/CICADA/GUPPiUsersGuide}} at the GBT) was configured for the settings shown in Table~\ref{tab:searchparams}. All data were taken in total-intensity, summed polarization mode. Once disks became full, they were shipped from Arecibo to Columbia University for data reduction. A summary of all observations is provided in Table~\ref{tab:survey}.

\begin{deluxetable*}{lcccc}
\tabletypesize{\scriptsize}
\centering
\tablecaption{\label{tab:searchparams} Observing Parameters for Four Radio Surveys of \fermi{}-LAT Sources}
\tablecolumns{5}
\tablehead{
\colhead{Parameter} &
\colhead{AO 327-MHz Survey} &
\colhead{GBT 350-MHz Survey\tablenotemark{a}} &
\colhead{GBT 820-MHz Survey\tablenotemark{b}} &
\colhead{Parkes 1390-MHz Survey\tablenotemark{c}}}
\startdata
Detection fraction\tablenotemark{d} & 8/34 (24\%); 5/6 & 13/50 (26\%); 3/13 & 3/25 (12\%); 1/3 & 11/56 (20\%); 2/11  \\
Center frequency (MHz) & 327 & 350 & 820 & 1390 \\
Bandwidth, $\Delta f$ (MHz) & 68.75\tablenotemark{e} & 100 & 200 & 256 \\
Number of channels & 2816\tablenotemark{f} & 4096 & 2048 & 512 \\
Sample time ($\mu$s) & 81.92 & 81.92 & 61.44 & 125 \\
Receiver temperature\tablenotemark{g}, $T_{\rm rec}$ (K) & 62 & 20 & 18 & 25 \\
Average sky temperature\tablenotemark{h}, $T_{\rm sky}$ (K) & 64 & 65 & 15 & 5 \\
Telescope gain, $G$ (K/Jy) & 10 & 2 & 2 & 0.735 \\
Effective threshold $S/N$ ($\beta\cdot S/N_{\rm min}$) & 10 & 10 & 10 & 12 \\
Integration time, $t_{\rm int}$ (minutes) & 15 & 32 & 45 & 60 \\
\enddata
\tablenotetext{a}{\cite{hrm+11}.}
\tablenotetext{b}{\cite{rrc+11}.}
\tablenotetext{c}{\cite{kcj+12,fc2015}.}
\tablenotetext{d}{Number of \emph{detected} MSPs divided by the total number of sources \emph{observed}; number of black widow plus redback systems \emph{discovered} divided by the total number of MSPs \emph{discovered}.}
\tablenotetext{e}{Recorded bandwidth: 100 MHz were sampled by PUPPI but we only recorded the section covering the receiver bandwidth.}
\tablenotetext{f}{Number of recorded channels; 4096 channels were sampled across the entire 100-MHz bandwidth.}
\tablenotetext{g}{Receiver temperature including spillover but excluding Galactic/CMB contribution. Values for Arecibo 327 MHz from NAIC\footnote{\url{http://www.naic.edu/~phil/cal327/327Calib.html}}. GBT values are from page 11 of the proposer's guide\footnote{\url{https://science.nrao.edu/facilities/gbt/proposing/GBTpg.pdf}}. See also \citet{lbs13} for 350 MHz. Parkes values are based on those from the users guide\footnote{\url{http://www.parkes.atnf.csiro.au/observing/documentation/user_guide/pks_ug_3.html\#Receivers-and-Correlators}}.}
\tablenotetext{h}{We calculated sky temperatures by scaling the \cite{hks81} 408-MHz map to each survey's observing frequency using a spectral index of $-2.6$ \citep{lmop87}. For AO 327, we list the average $T_{\rm sky}$ for each of the 34 target locations. For GBT 350 and Parkes, we calculated the average temperature at an evenly spaced grid of points encompassing the search regions. For GBT 820 we averaged the \citet{rrc+11} values for their searches excluding the Galactic plane.}
\end{deluxetable*}

Data were analyzed using the software package \verb|PRESTO| \citep[][]{ran01}. The data reduction process began with the detection and masking of significant radio frequency interference in the data. Dedispersion occurred up to a specified dispersion measure (DM), which we chose to be twice the maximum line-of-sight value given by the NE2001 model \citep{cl02}. \verb|PRESTO| can perform searches over spin period variations caused by orbital motion, searching over both period and period derivative. The extent of the acceleration search is specified by the \verb|zmax| parameter, chosen to be 200 in our case. This means that linear pulsar spin frequency ($f_0$) drifts of up to 200 bins were searched in the highest harmonic, which in our analysis was the eighth \citep{rem02}. If $t_{\rm int}$ is the total integration time and $a_{\rm max}$ is the maximum acceleration probed, $z_{\rm max} = a_{\rm max} t_{\rm int}^2 f_0/c$ \citep{rgh2000}.

\subsection{Sensitivity}\label{sec:AOsens}
\begin{figure}
\begin{center}
\includegraphics[width=\linewidth,trim={80 0 80 0},clip]{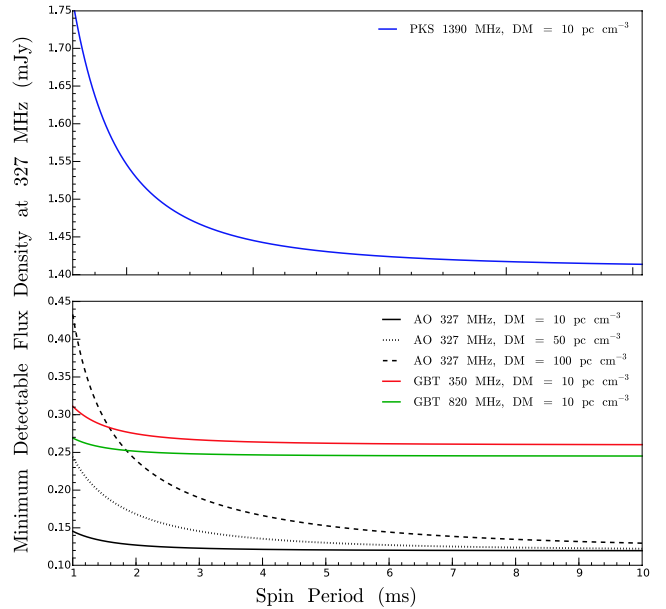}
\end{center}
\caption{\label{fig:sens} Sensitivity of four radio pulsar searches of \fermi{}-LAT sources as a function of spin period for a pulsar with 10\% duty cycle. All surveys have been scaled to 327 MHz using a spectral index of $-1.7$. The surveys presented here, in addition to this work, are: GBT 350 MHz \citep{hrm+11}, GBT 820 MHz \citep{rrc+11}, and Parkes 1390 MHz \citep{fc2015}. See Table~\ref{tab:searchparams} for details.}
\end{figure}

Figure~\ref{fig:sens} shows the minimum flux density detectable by our Arecibo searches for a range of MSP spin periods and DMs, as determined by the radiometer equation for pulsars \cite[][Appendix 1.4]{lk05}:
\begin{equation}\label{eq:sens}
S_{\rm min} = \beta\frac{S/N_{\rm min}\textnormal{ }(T_{\rm rec}+T_{\rm sky})}{G\sqrt{n_{\rm p}t_{\rm int}\Delta f}}\sqrt{\frac{W}{P-W}},
\end{equation}
where $S_{\rm min}$ is the minimum detectable flux density, $\beta$ is a normalization factor including corrections for, e.g., system digitization losses, $S/N_{\rm min}$ is the threshold pulsar signal-to-noise ratio, $T_{\rm rec}$ is the receiver temperature (including contributions from the CMB and spillover), $T_{\rm sky}$ is the sky temperature, $W$ is the effective pulse width (we assume the intrinsic pulse width to be $P/10$), $P$ is the pulsar period, $G$ is the telescope gain, $n_{\rm p}$ is the number of polarizations recorded (always 2 for the searches discussed here), $t_{\rm int}$ is the integration time, and $\Delta f$ is the effective bandwidth. Relevant parameters for the Arecibo survey are shown in Table~\ref{tab:searchparams} under ``AO 327''.  Arecibo's system equivalent flux density (SEFD) degrades for zenith angles exceeding 15 degrees\footnote{Detailed measurements for gain and system temperature of the 327-MHz Gregorian receiver at Arecibo can be found at \url{http://www.naic.edu/~astro/RXstatus/327/327greg.shtml}.}, but this had little impact for most of our searches. Overall our survey had an average $\mbox{SEFD} = 13$\,Jy. The sensitivity of the Arecibo survey in the context of other \fermi-LAT searches is discussed in Section~\ref{sec:discussion}.

\section{Results}\label{sec:results}
\begin{figure*}
\centering
\begin{tabular}{cccc}
(a) J0251+26 & (b) J1048+2339 & (c) J1805+06 & (d) J1824+10 \\
\includegraphics[width=0.15\textwidth,trim={80 0 1500 150},clip]{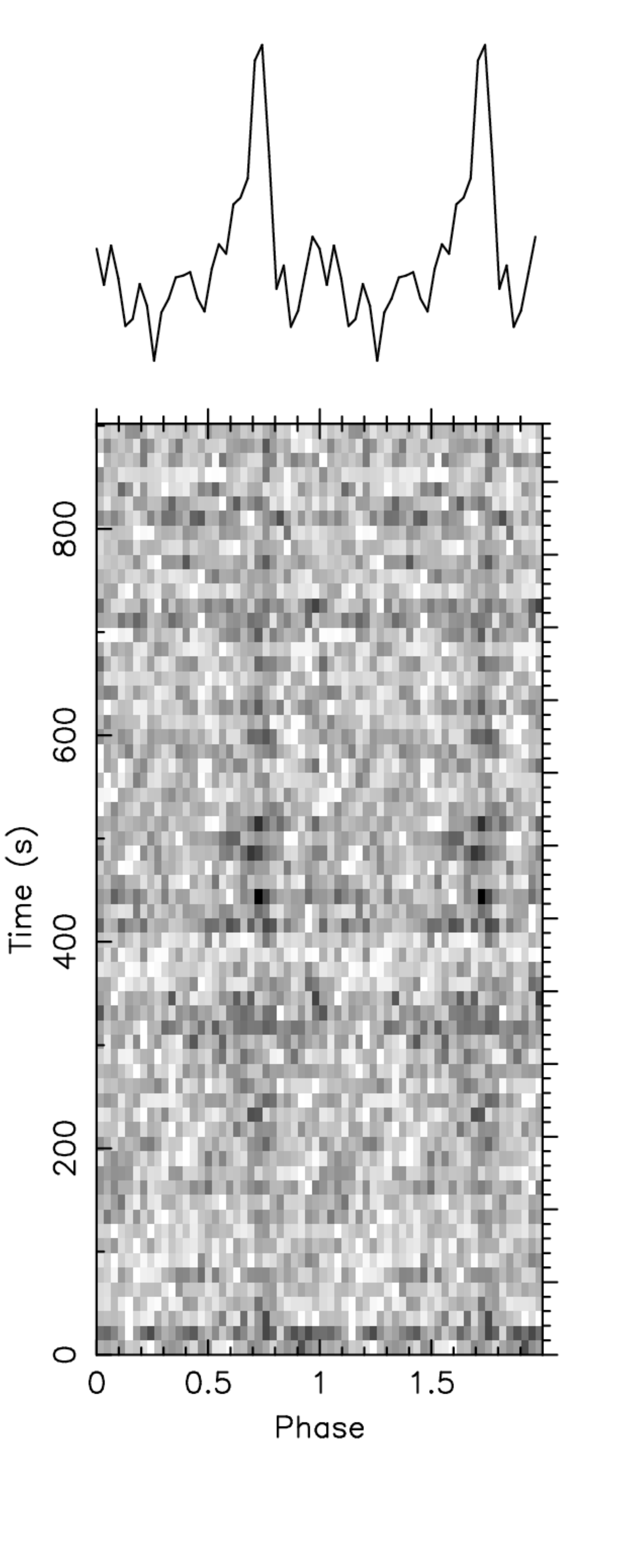} &   \includegraphics[width=0.15\textwidth,trim={80 0 1500 150},clip]{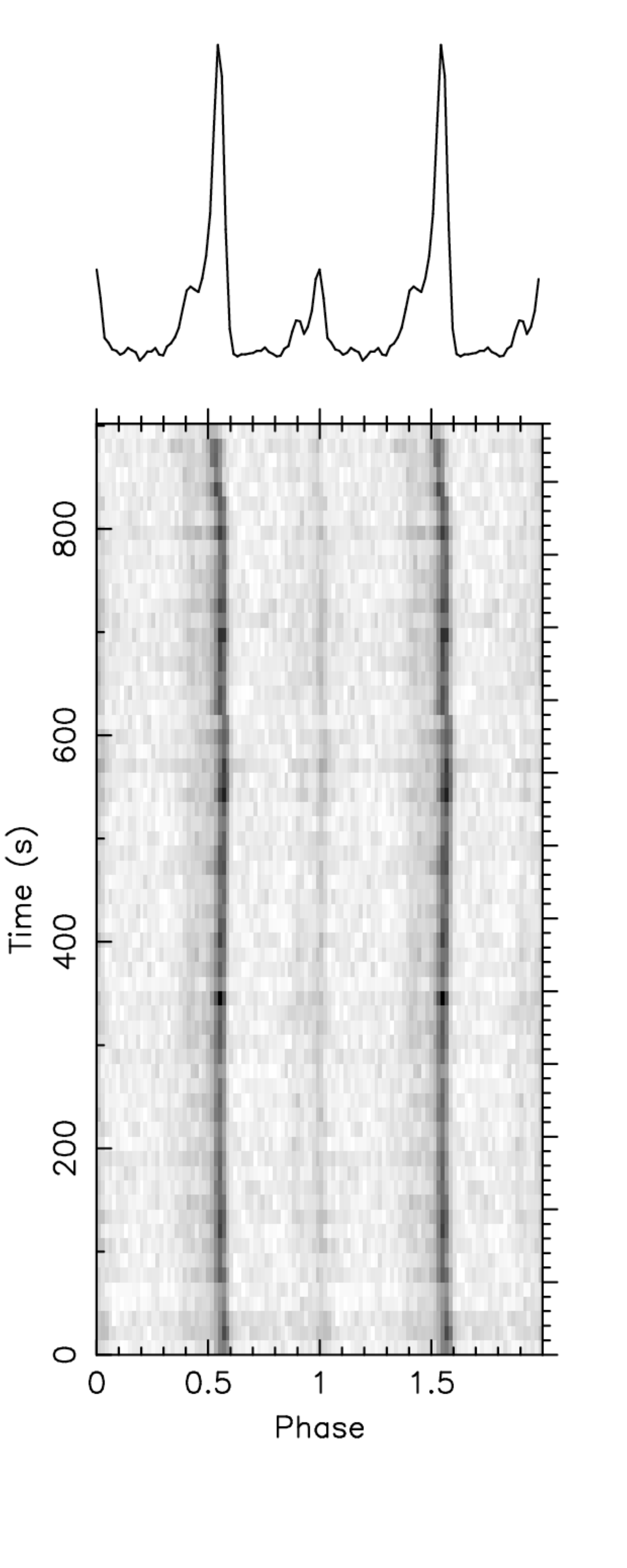} & \includegraphics[width=0.15\textwidth,trim={80 0 1500 150},clip]{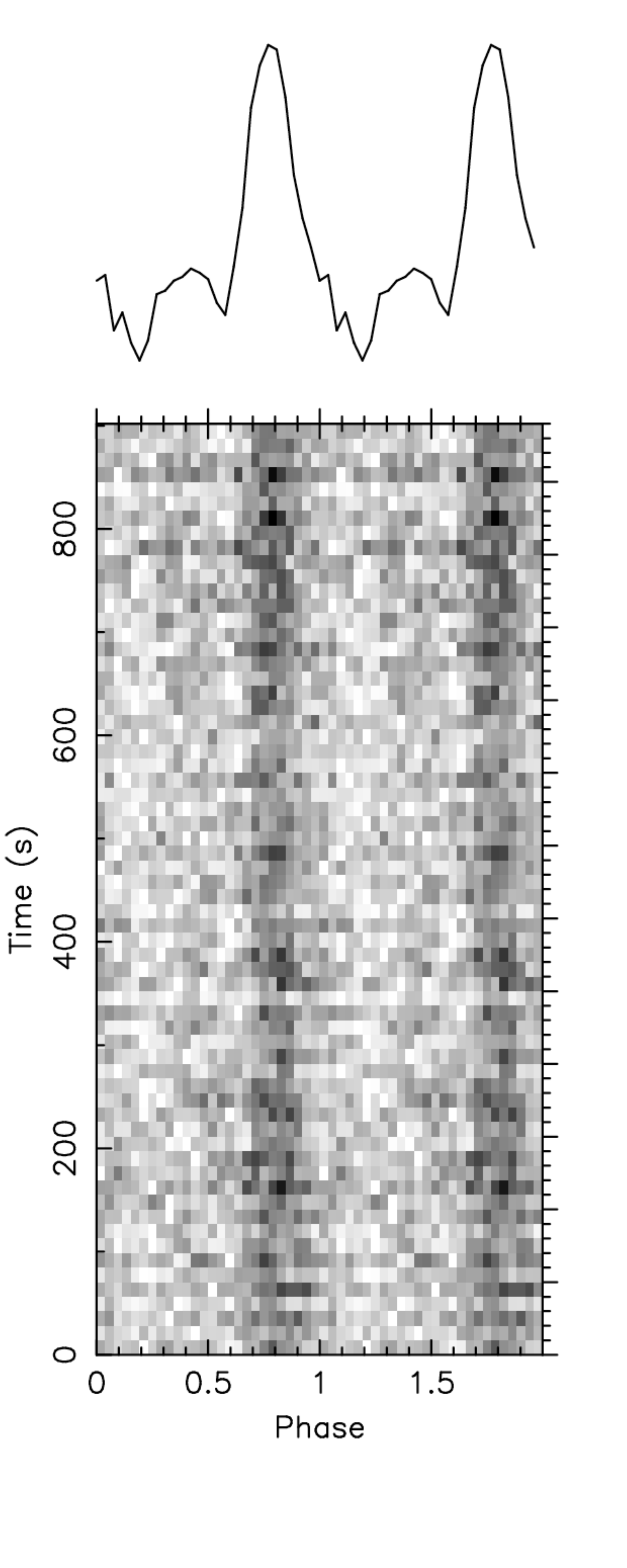} & \includegraphics[width=0.15\textwidth,trim={80 0 1500 150},clip]{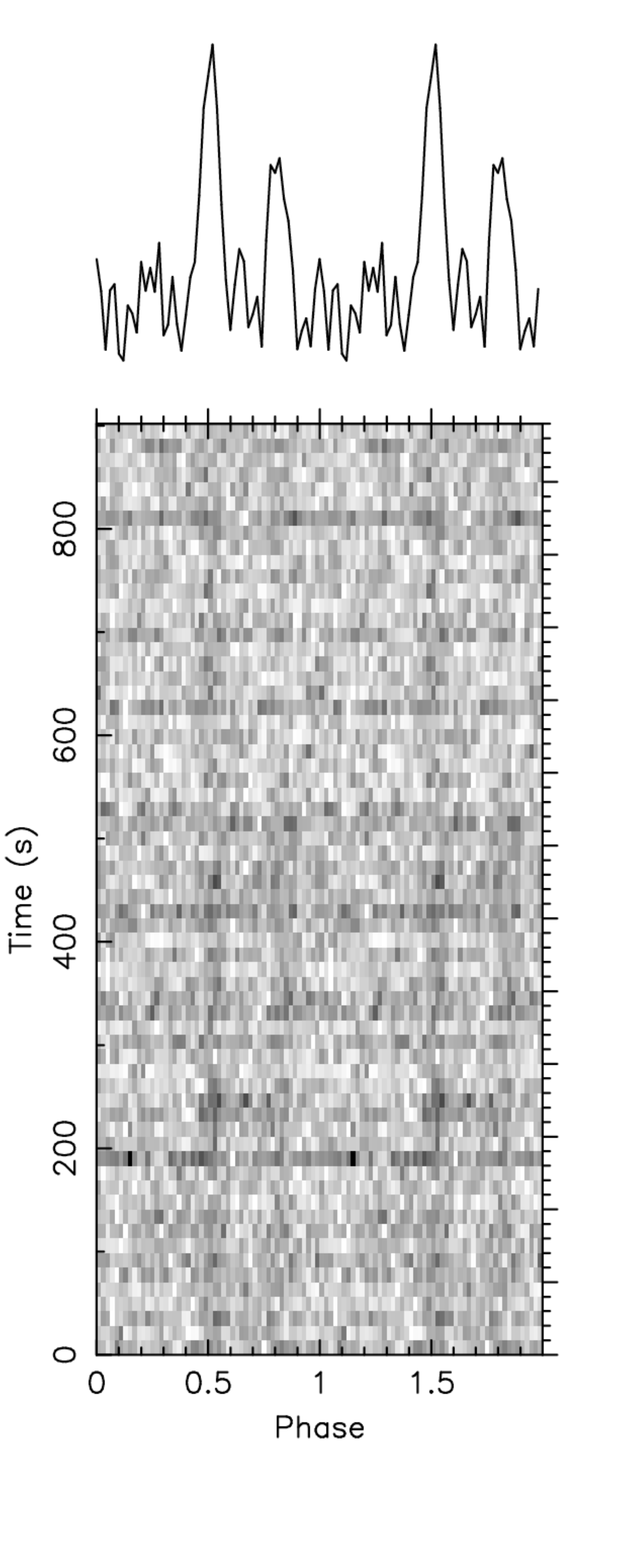} \\ (e) J1909+21 & (f) J2052+1218 & (g) J2052+1218 & (h) J2052+1218\\
&  (2013 Jun 25) & (2013 Jul 04) & (2013 Sep 12)\\
\includegraphics[width=0.15\textwidth,trim={80 0 1500 150},clip]{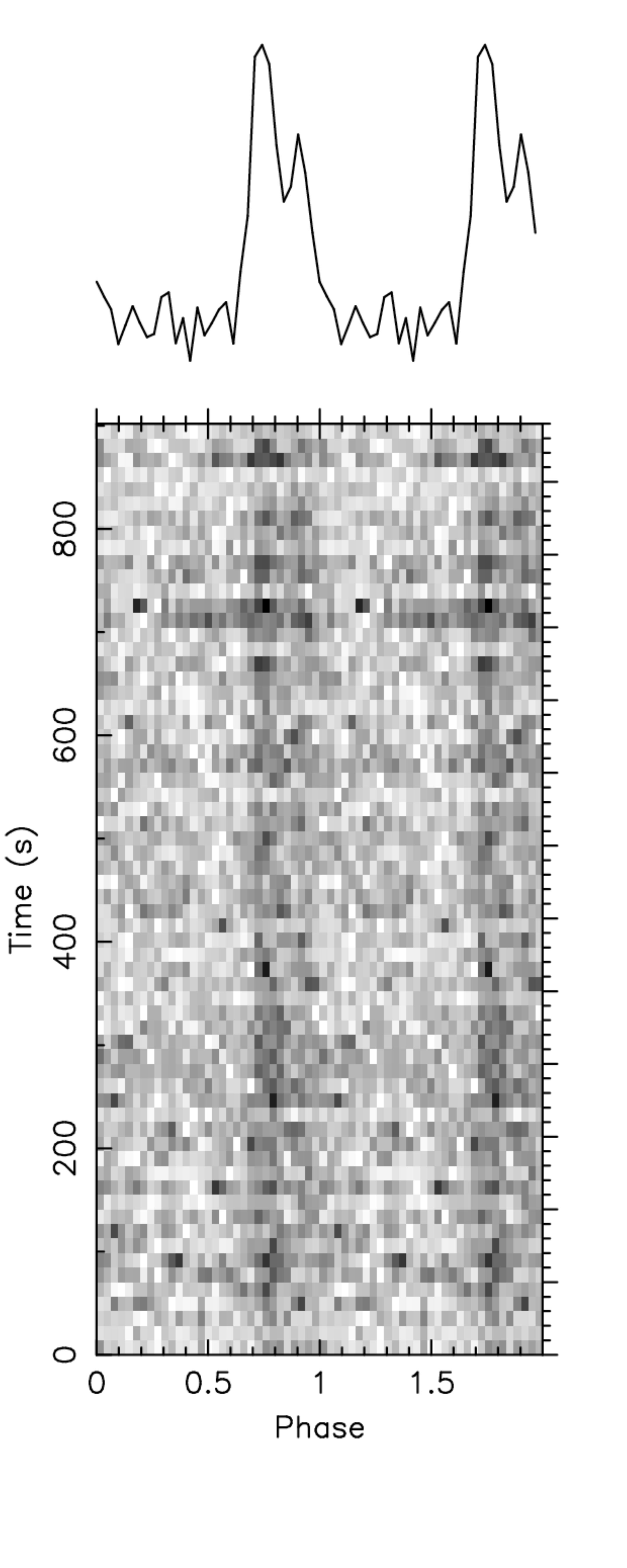} &   \includegraphics[width=0.15\textwidth,trim={80 0 1500 150},clip]{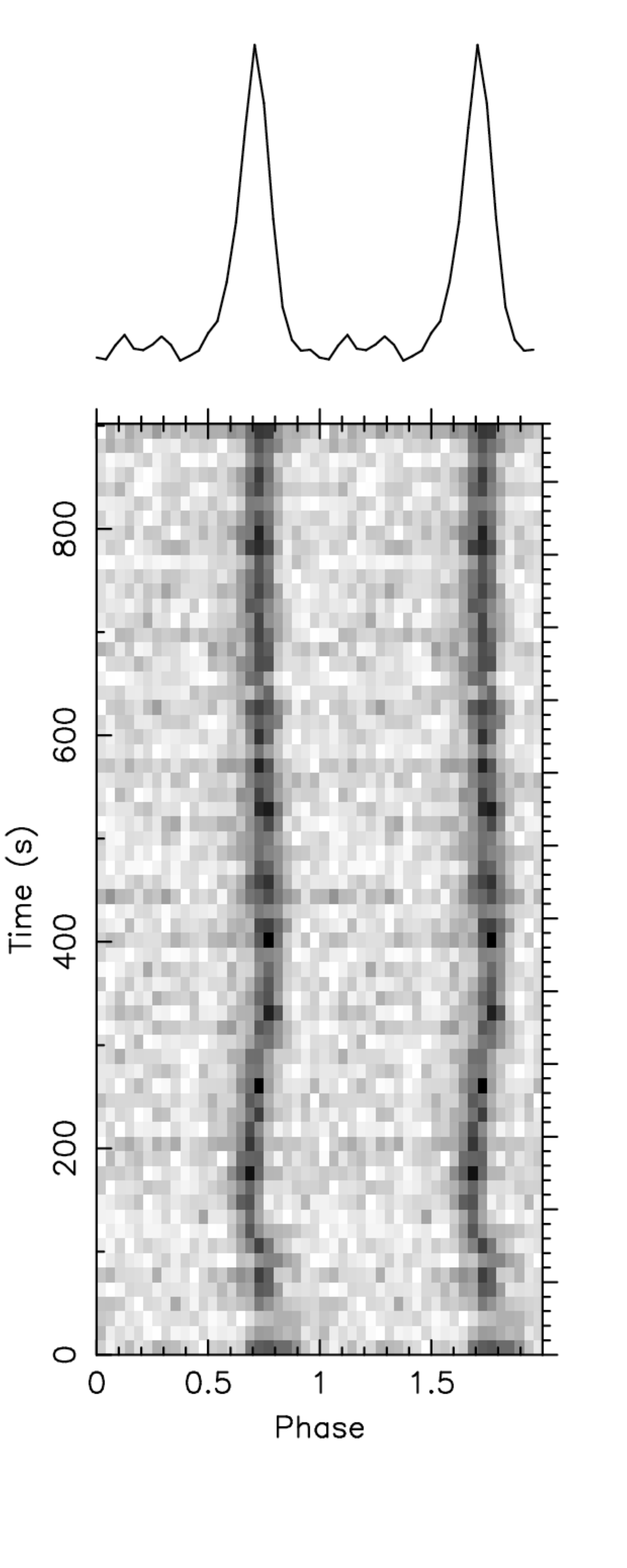} & \includegraphics[width=0.15\textwidth,trim={80 0 1500 150},clip]{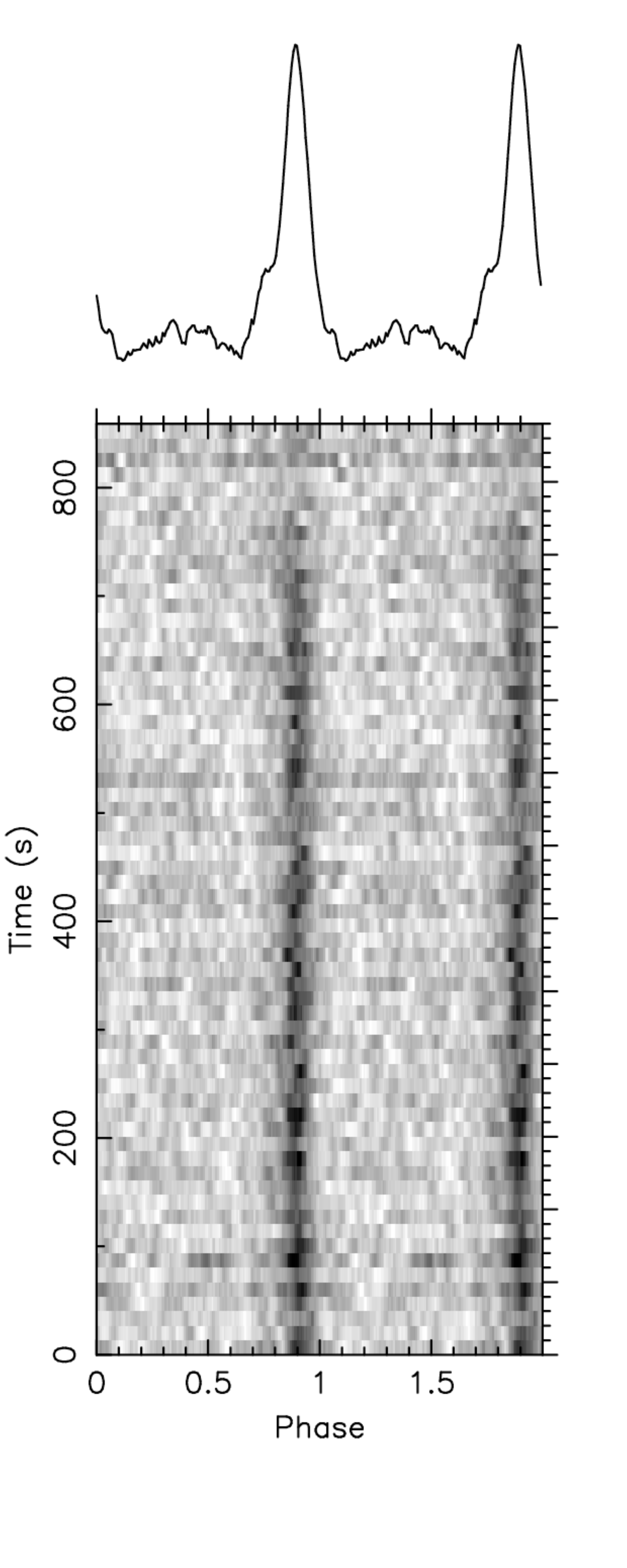} &   \includegraphics[width=0.15\textwidth,trim={80 0 1500 150},clip]{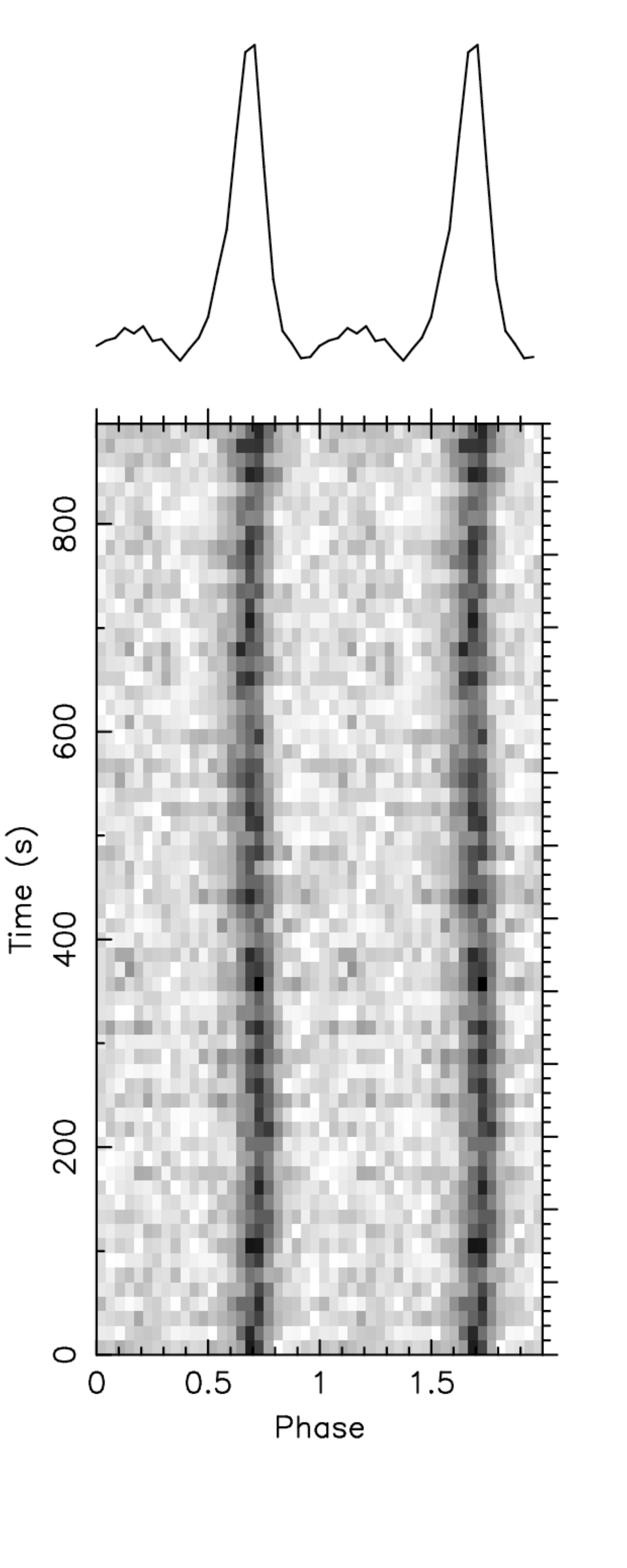}\\
\end{tabular}
\caption{\label{fig:pulses}The best detections from the search observations of six new MSPs, folded modulo the period and period derivative returned by the software (two rotations are shown). The three search observations of the eclipsing \psrf{} show some effects related to likely eclipse egress (panel f) and ingress (g).}
\end{figure*}

In the 34 sources searched, we discovered six new MSPs (see Table~\ref{tab:results}). Pulse profiles from the discovery observations are shown in Figure~\ref{fig:pulses}. The rotation periods range between 1.99 and 4.66 ms, and their DMs span 17--65\,pc $\textnormal{cm}^{-3}$. Orbital solutions have been obtained for all new discoveries from initial timing observations. However, phase-connected timing solutions (including precise positions, period derivatives, spin-down luminosities, and a study of the gamma-ray properties of the coincident gamma-ray sources)  are not yet available for most of the MSPs, and will be presented elsewhere. A study of the redback \psrb{} is presented in \cite{den15}.

\begin{deluxetable*}{lcccccccc}
\tabletypesize{\scriptsize}
\centering
\tablewidth{0pt}
\tablecaption{\label{tab:results} Pulsars discovered in Arecibo Searches of \fermi-LAT Sources}
\tablecolumns{9}
\tablehead{
\colhead{Name\tablenotemark{a}} &
\colhead{$P$} &
\colhead{DM} &
\colhead{Distance\tablenotemark{b}} &
\colhead{$P_{\textnormal{orbit}}$} &
\colhead{Minimum Companion} &
\colhead{Type\tablenotemark{d}} &
\colhead{Eclipses?} &
\colhead{Discovery Flux Densities}\\
\colhead{} &
\colhead{(ms)} &
\colhead{(pc cm$^{-3}$)} &
\colhead{(kpc)} &
\colhead{(hr)} &
\colhead{Mass\tablenotemark{c} (\msun)} &
\colhead{} &
\colhead{} &
\colhead{(mJy)}}
\startdata
J0251+26 & 2.54 & 20 & 0.8 & 4.9 & 0.024 & BW & Yes & 0.3, 0.3\\
J1048+2339 & 4.66 & 17 & 0.7 & 6.0 & 0.30 & RB & Yes & 2.4\\
J1805+06 & 2.13 & 65 & 2.5 & 8.1 & 0.023 & BW & No\tablenotemark{e} & 1.1, 1.5\\
J1824+10 & 4.07 & 60 & 2.5 & 1980.0 & 0.26 & NSWD & No & 0.09, 0.15\\
J1909+21 & 2.56 & 62 & 3.2 & 3.5 & 0.055 & RB & Yes & 0.6\\
J2052+1218 & 1.99 & 42 & 2.4 & 2.6 & 0.033 & BW & Yes & 1.3, 1.0, 1.2\\
\enddata
\tablenotetext{a}{Names with four digits of declination have been given only to MSPs with phase-connected timing solutions.}
\tablenotetext{b}{From the NE2001 model \citep{cl02}.}
\tablenotetext{c}{Assuming a pulsar mass of 1.35 \msun{} \citep{oz12}.}
\tablenotetext{d}{BW = black widow, RB = redback, NSWD = neutron star-white dwarf.}
\tablenotetext{e}{\psrc{} has not yet shown any eclipsing behavior; however, there is a gap in orbital coverage at phases 0.2--0.27, so eclipses cannot be ruled out.}
\end{deluxetable*}

Five of the new MSPs are neutron stars with short orbital periods. Three are ``black widows'', in which much of the companion mass has been stripped away or accreted by the pulsar, leaving a (partially degenerate) companion with mass $\ll 0.1 \textnormal{ M}_{\odot}$. The remaining two short-orbit systems are ``redbacks'', where the pulsar is frequently eclipsed by outflows from a non-degenerate companion with mass $\ga 0.1 \textnormal{ M}_{\odot}$. The final MSP is a more classical neutron star-white dwarf binary. For a diagram of orbital period vs.\ companion mass for such highly accelerated systems, see \cite{rob12}.

\begin{figure}
\centering
\includegraphics[width=\linewidth]{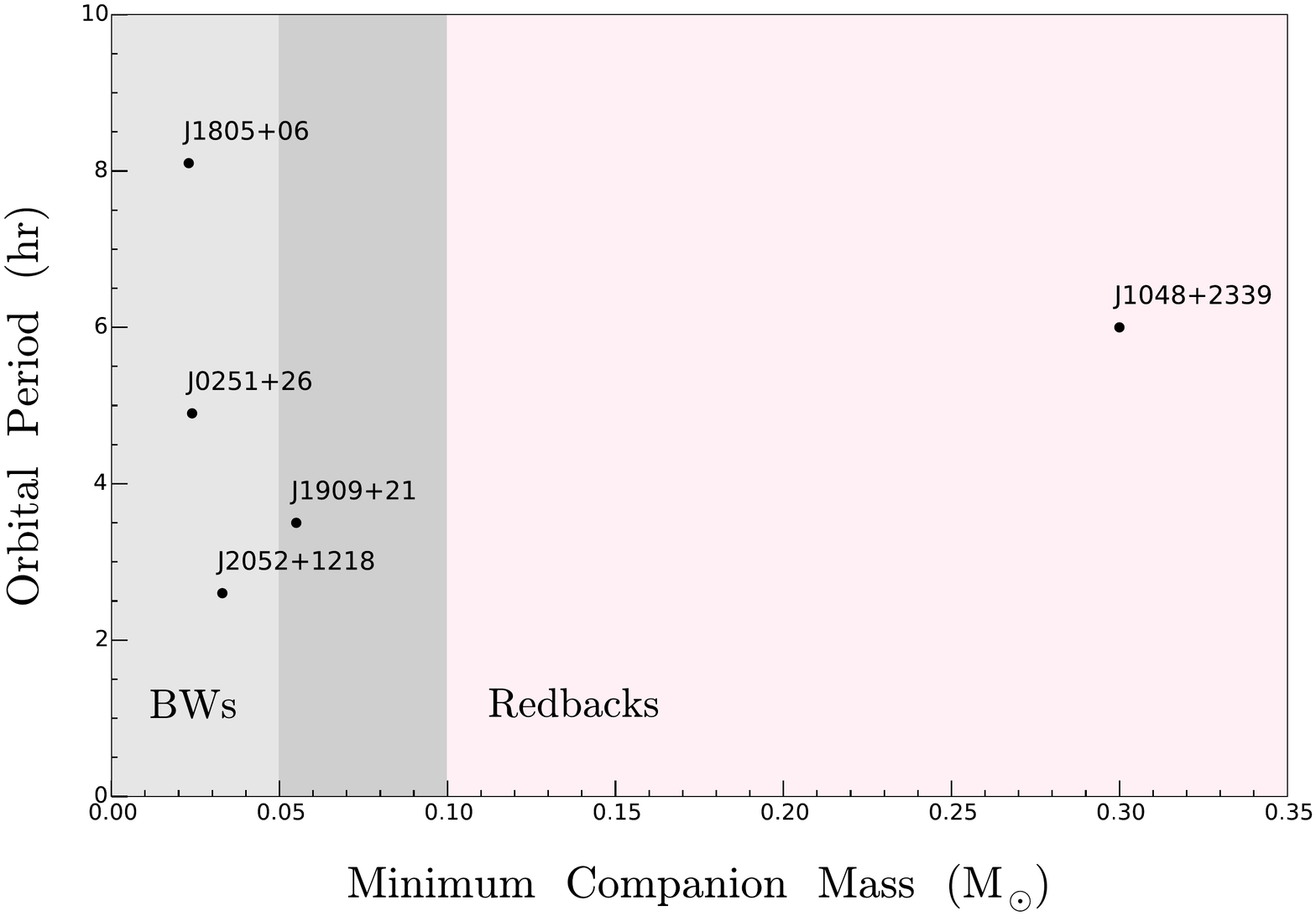}
\caption{The new short-orbit MSPs from this work are presented in an orbital period vs. minimum companion mass plot. MSPs in the light grey area (leftmost block) are black widows, the one in the pink area (rightmost block) is a redback, and \psre{} is intermediate between the two, but classified as a redback (see Section~\ref{sec:results}).} \label{fig:orbmass}
\end{figure}

Figure~\ref{fig:orbmass} shows the distribution of orbital periods vs.\ minimum companion masses for five of the new MSPs presented in this work. Minimum companion masses are calculated using Keplarian parameters derived from orbital timing solutions. Note that \psre{} is classified as a redback, despite its minimum companion mass being less than 0.1 M$_{\odot}$. This is firstly because the 0.055 M$_{\odot}$ value is the \emph{minimum} companion mass, and it is larger than any known black widow minimum companion mass; secondly, its eclipses last for approximately half of the orbit, which is characteristic of a redback system with a dense circumstellar environment.

\psrc{} is in a black widow system with an orbital period of 8.1 hr. The approximate position of 3FGL J1805.9+0614 was observed in 2009 at the Robert C. Byrd Green Bank Telescope (GBT) at 350 MHz (M. Roberts, private communication); however, only a quick search of the first five minutes of data was performed and the pulsar was not detected. Searching the full data set following our discovery at Arecibo, the MSP is clearly detected. Ransom et al.\ also observed this source twice with the GBT at 820 MHz, but the MSP did not show up in a preliminary analysis of the first observation and the second dataset was not searched. Using the known DM and approximate period from our Arecibo detections reveals the pulsar in both GBT datasets.

\psrf{} is an intriguing system due to the pulsar's very fast rotation (1.99 ms) and its short binary period (2.8\,hr).  Even after searching over acceleration, residual drifts in phase vs.\ time can be seen in this and other black widow and redback systems (see Figure~\ref{fig:pulses}, especially (b) and (h)).

We searched the sources containing PSRs~J1921+01 and J2042+02 and detected the MSPs, unaware that they had already been discovered at the GBT. These will be published in a forthcoming paper detailing \fermi{}-LAT searches at the GBT (S.\ Sanpa-Arsa et al.\ 2016, in preparation).

The six new findings mark the first \fermi{} MSPs discovered using the Arecibo telescope and broke the 50-pulsar threshold for total LAT-guided radio MSP discoveries (which as of 2015 December stands at 69).

\section{Discussion}\label{sec:discussion}
\subsection{Possible Candidates for Re-Observation}
In searching 34 unidentified \fermi{}-LAT gamma-ray sources at Arecibo, we detected 8 MSPs, for a 24\% success rate. This is in line with the success rate for LAT-guided radio surveys at the GBT and Parkes \citep[][see Table~\ref{tab:searchparams}]{hrm+11,rrc+11,fc2015}. While we find this to be a satisfying result, it is possible that some remaining sources in Table~\ref{tab:survey2} could still be pulsars. Seventeen of the 26 sources currently without a known pulsar counterpart are spectrally consistent with pulsars (denoted by a ranking of 1, 2, or 3 in the ``Spectrum Notes'' column) and have no known AGN association. Sources ranked 1 or 2 are very likely to be pulsars, while rank-3 sources lack definitive evidence to suggest they are not pulsars. An inability to make a detection does not preclude the presence of an MSP; rather, it may be due to a pulsar's faintness, eclipses, scintillation, or extreme orbital or spin parameters. The large number of black widow and redback systems that have been discovered in \fermi{}-LAT sources make eclipses a distinct possibility for this collection of candidates.  For example, both PSRs~J1048+2339 and J1909+21 were only detected in the second of two search observations, owing to eclipses (Table~\ref{tab:survey}). Additional observations of the 17 remaining ``good'' sources in Table~\ref{tab:survey2} may result in the detection of new MSPs.

\begin{deluxetable*}{l@{\hspace{-0.1cm}}rcrrrcr}
\tablewidth{0pt}
\tablecaption{\label{tab:survey2} Arecibo Searches: Unidentified Gamma-Ray Source Information from 3FGL Catalog}
\tablecolumns{8}
\tablehead{
\colhead{3FGL name\tablenotemark{a}}       &
\colhead{$r95$\tablenotemark{b}}           &
\colhead{Class\tablenotemark{c}}           &
\colhead{Sig\tablenotemark{d}}             &
\colhead{Curve\tablenotemark{e}}           &
\colhead{Var\tablenotemark{f}}             &
\colhead{Spectrum\tablenotemark{g}}        &
\colhead{$N_{\rm obs}$\tablenotemark{h}}   \\
\colhead{}                              &
\colhead{(deg)}                         &
\colhead{}                              &
\colhead{($\sigma$)}                    &
\colhead{($\sigma$)}                    &
\colhead{}                              &
\colhead{Notes}                         &
\colhead{}
}
\startdata
\sout{J0103.7+1323}  & 0.08 & bcu  & 7.1 & 2.4 & 53 & 3 lh  &   3 \\ 
\sout{J0134.5+2638}  & 0.06 & bcu  & 12.0 & 3.0 & 58 & 4 lh  &   3 \\ 
\sout{J0232.9+2606}  & 0.09 & bcu  & 4.3 & 1.6 & 34 & 3 h  &   3 \\ 
\textbf{J0251.1+2603}  & 0.11 & psr  & 7.9 & 3.4 & 36 & 2 cp  &   2 \\ 
J0318.1+0252  & 0.09 & \nodata  & 12.8 & 5.7 & 50 & 1 Pc  &   3 \\ 
J0330.6+0437  & 0.11 & \nodata  & 7.4 & 2.4 & 62 & 2 cd  &   3 \\
J0342.3+3148c  & 0.10 & \nodata  & 6.8 & 2.5 & 48 & 3 ld  &   3 \\ 
J0421.6+1950  & 0.12 & \nodata  & 6.1 & 1.9 & 47 & 2 ld  &   3 \\ 
J0517.1+2628c  & 0.12 & \nodata  & 7.0 & 1.6 & 50 & 3 ld  &   5 \\ 
\sout{J0539.8+1434}  & 0.08 & fsrq  & 7.6 & 3.1 & 299 & 5 lVd  &   2 \\ 
\textbf{J1048.6+2338}  & 0.12 & bll  & 9.0 & 2.5 & 50 & 3 LD  &   2 \\ 
J1049.7+1548  & 0.09 & \nodata  & 7.6 & 1.1 & 58 & 3 lh  &   2 \\ 
\sout{J1200.4+0202}  & 0.07 & \nodata  & 8.8 & 1.6 & 56 & 4 Lh  &   3 \\ 
J1225.9+2953  & 0.05 & \nodata  & 17.4 & 5.2 & 58 & 1 Cp  &   3 \\ 
P7R4 J1250+3118\tablenotemark{i}  &   \nodata  &   \nodata  &   \nodata  &   \nodata  &   \nodata  &   \nodata  &   3 \\
J1309.0+0347  & 0.15 & \nodata  & 4.1 & 2.7 & 43 & 2 ?l  &   6 \\ 
\sout{J1322.3+0839}  & 0.12 & bcu  & 7.8 & 0.4 & 74 & 3 ld  &   3 \\ 
J1601.9+2306  & 0.11 & \nodata  & 7.8 & 4.3 & 47 & 2 P  &   3 \\ 
J1627.8+3217  & 0.07 & \nodata  & 10.2 & 3.8 & 33 & 2 C  &   2 \\ 
\sout{J1704.1+1234}  & 0.07 & \nodata  & 9.4 & 0.5 & 47 & 4 LD  &   3 \\ 
J1720.7+0711  & 0.09 & \nodata  & 9.6 & 1.6 & 44 & 3 cD  &   3 \\ 
\textbf{J1805.9+0614}  & 0.09 & psr  & 9.3 & 4.3 & 40 & 1 CP  &   2 \\ 
\textbf{J1824.0+1017}  & 0.09 & psr  & 6.7 & 3.5 & 48 & 2 lc  &   2 \\
J1827.7+1141  & 0.10 & \nodata  & 6.5 & 3.8 & 39 & 2 lc  &   2 \\ 
J1829.2+3229  & 0.15 & \nodata  & 5.7 & 3.5 & 49 & 2 ld  &   3 \\ 
J1842.2+2742  & 0.08 & \nodata  & 8.3 & 2.4 & 39 & 2 c  &   3 \\ 
\textbf{P7R4 J1909+2102\tablenotemark{i}}  &   \nodata  &   \nodata  &   \nodata  &   \nodata  &   \nodata  &   \nodata  &   2 \\
\textbf{J1921.2+0136}  & 0.10 & psr  & 9.3 & 1.7 & 35 & 2 cD  &   4 \\ 
J2026.3+1430  & 0.09 & \nodata  & 7.1 & 2.1 & 49 & 2 c  &   3 \\ 
\textbf{J2042.1+0247}  & 0.14 & PSR  & 7.5 & 4.8 & 28 & 1 CP  &   2 \\ 
\textbf{J2052.7+1217}  & 0.11 & psr  & 7.2 & 1.6 & 39 & 3 lD  &   3 \\ 
\sout{J2108.0+3654}  & 0.06 & bcu  & 6.4 & 1.5 & 40 & 4 Hl  &   3 \\ 
J2212.5+0703  & 0.10 & \nodata  & 14.3 & 4.2 & 57 & 1 Cp  &   6 \\ 
\sout{J2352.0+1752}  & 0.07 & bll  & 8.9 & 1.4 & 56 & 4 Hl  &   4 \\
\enddata
\tablecomments{3FGL source properties are from the \citet{aaa+15} catalog.}
\tablenotetext{a}{The six new Arecibo MSPs and two MSPs independently discovered at the GBT (S.\ Sanpa-Arsa et al.\ 2016, in preparation) are shown in bold.}
\tablenotetext{b}{3FGL source error circle radius at the 95\% confidence level.}
\tablenotetext{c}{3FGL pipeline classification scheme. PSR and psr are pulsars with and without LAT pulsations, respectively. The bll designation signifies a BL Lac object, bcu is an unclassified blazar, and fsrq is a flat-spectrum radio quasar.}
\tablenotetext{d}{3FGL source significance.}
\tablenotetext{e}{Curvature significance for 3FGL source spectrum when fit to a log-parabolic model.}
\tablenotetext{f}{Variability index for source, where an index $>73$ denotes variability at $>99\%$ confidence level.}
\tablenotetext{g}{For a full description of \emph{Spectrum Notes}, see \cite{fc2015}. The first number in this scheme is a rating of how likely the source is to be a pulsar. A ``1'' means it is very likely, while sources with a ``4'' or ``5'' rating (or ``3'' with a possible AGN association) have been crossed off the list and will not be reobserved because they are unlikely to be pulsars. The source characteristics, on which the rating is based, are obtained from inspection of a source's spectral energy distribution.}
\tablenotetext{h}{Number of times each 3FGL source was observed (from Table~\ref{tab:survey}).}
\tablenotetext{i}{Not included in the 3FGL catalog. P7R4 designators refer to unpublished source lists. }
\end{deluxetable*}

\subsection{Uncertain Gamma-Ray Associations}
Two new MSPs, PSRs~J1048+2339 and J1909+21, may have been ``lucky'' discoveries within the error circle of a gamma-ray source, but not necessarily associated with that source. In 3FGL, J1048.6+2338 is listed as being possibly associated with a BL Lacertae-type blazar. Blazar associations are generally spatial, and accidental coincidence is a common cause for reclassification of non-variable sources. Until it is possible to fold the gamma-ray photons modulo the parameters obtained with a radio timing solution, it will remain unclear whether the \fermi{}-LAT source is an MSP or possibly a blazar. \psre{} is not associated with a nearby 3FGL source. We selected it for observation because in a preliminary source list internal to the LAT collaboration there appeared to be a promising source. As for all the MSPs we have discovered, we will know whether this one is associated with a LAT source once we have rotational ephemerides and can fold the gamma-ray photons.

\subsection{Sensitivity in the Context of Other LAT Radio Surveys}\label{sec:sens}
If no break exists in the $\log N - \log S$ distribution of MSPs, one might expect that an increase in sensitivity would yield higher survey success rates. Instead, our discovery rates were comparable to those of other lower-sensitivity LAT-guided MSP surveys. However, we have based our target list on the 3FGL catalog, while previous surveys have been based largely on earlier catalogs, and newer {\em Fermi}-LAT catalogs include weaker, less well characterized sources that are more difficult to classify spectrally.  In addition, some relatively bright MSPs, particularly those not subject to large accelerations, would have already been discovered in previous Arecibo ``all-sky'' surveys.

Figure~\ref{fig:sens} presents minimum detectable flux densities for four radio searches of \fermi{}-LAT sources, including our Arecibo work.  Parameters for each of the surveys are provided in Table~\ref{tab:searchparams}. Each of the sensitivity curves has been scaled to 327 MHz using an assumed MSP spectral index of $\alpha = -1.7$ \citep{sto14}. As an example of their relative power, for spin period $P = 1.8$\,ms the Arecibo searches are as sensitive at $\mbox{DM} = 100$\,pc\,cm$^{-3}$ as the GBT surveys are for $\mbox{DM} = 10$\,pc\,cm$^{-3}$. For identical low DMs, the Arecibo surveys are about twice as sensitive as the GBT searches. In other words, integration time at the GBT would have to be quadrupled to reach comparable raw sensitivities to Arecibo --- but such an increase in integration time would have deleterious consequences for the detectability of compact binaries.

We list the radio flux densities for all discovery observations in Table~\ref{tab:results} (these were obtained from an application of the radiometer equation and we estimate they have $\approx 25\%$ uncertainty). We see by comparison to Figure~\ref{fig:sens} that PSRs~J0251+26 and J1824+10 could only have been discovered with Arecibo. Parkes could only have detected \psrb{}. This only considers raw telescope sensitivity; it does not take into account sensitivity to high acceleration (discussed later), which further emphasizes the utility of large telescopes. For a discussion of selection effects related to interstellar scintillation and eclipses, see \cite{fc2015}.

Why has the Arecibo survey turned up such a large proportion (5/6) of highly accelerated interacting binaries, compared to the fractions found in other \fermi{}-LAT surveys? Though small-number statistics is a possible explanation, the result can likely be attributed to the Arecibo telescope's very large gain, coupled with the relatively short integration times used, and the multiple-observation strategy used to search each good target. An integration time of just 15 minutes at Arecibo yields a minimum detectable flux density that is substantially lower than the longer integrations elsewhere (see Figure~\ref{fig:sens}).

The population of \fermi{}-LAT MSPs contains a disproportionately large number of interacting binary systems for reasons that are currently poorly understood. For a time, it was thought that a tendency for intrabinary shocks to produce high-energy radiation could bias \fermi-LAT searches towards discovering these systems \citep[e.g.,][]{rap+12}. More recent analyses, however, have found little evidence to support this claim \citep{tj15}. The bias is likely due in part to previous surveys' biases against finding binaries due to eclipses and acceleration.

The use of modern acceleration search techniques (as implemented within \verb|PRESTO| in our case; Ransom 2001)\nocite{ran01} was essential for the detection of the five compact MSP systems.  Both \cite{jk91} and \cite{blw13} have explored the detectability of binary pulsar systems, the latter having expanded the former's work to include eccentric binaries.  \cite{jk91} provide a quantitative measure of the loss of power due to acceleration by way of an ``efficiency factor'', $\gamma_m$.  Squaring this value gives a ratio of the power in the m$^{\rm th}$ harmonic, which includes degradation due to acceleration $a$ and jerk $\dot a$, to the power that would be present were the acceleration zero. Three such $\gamma_m$ terms were reformulated in \cite{blw13}. The first, $\gamma_{1m}^2$, describes the ratio that would be found in a ``standard'' pulsar search in which acceleration is not searched over. The term $\gamma_{2m}^2$ describes the power that would be recovered in a constant acceleration search (like the ones we performed), and will be employed here. It is formulated as follows:
\begin{equation}
\gamma_{2m} = \frac{1}{t_{\rm int}}\left|\int_0^{t_{\rm int}}{exp\left[\frac{im\omega_p}{c}\left(\left(\int_0^t{v_l dt}\right)-\alpha_a t^2-\alpha_v t\right)\right]dt}\right|,
\end{equation}
where $t_{\rm int}$ is the integration time of the observation, $v_l$ is the pulsar's line-of-sight velocity, and $\omega_{p}$ is its angular spin frequency.  A modern search algorithm yields values of acceleration $\alpha_a$ and velocity $\alpha_v$ that maximize $\gamma_{2m}$. Here, $\gamma_{2m}^2=1$ for a system with constant acceleration.  The final term, $\gamma_{3m}^2$, describes the power ratio recovered in a search over velocity, acceleration, and jerk. Such search algorithms are currently being developed, but have not yet been implemented.

Using software provided by \cite{blw13}\footnote{\url{http://psrpop.phys.wvu.edu/binary}}, we calculated values of $\gamma^2_{2m}$ for \psrf{} during a 15-minute integration, such as at Arecibo. We then recalculated these values using the integration times for surveys at the GBT and Parkes to compare the detectability of this fast-spinning, highly accelerated binary pulsar by the four different surveys. Results are given in Table~\ref{tab:gamma}.

\begin{deluxetable}{ccccc}
\centering
\tablecaption{\label{tab:gamma} Values of $\gamma^2_{2m}$ for \psrf{} as a Function of Integration Time}
\tablewidth{0pt}
\tablecolumns{5}
\tablehead{
\colhead{m (harmonic \#)} &
\colhead{15 min} &
\colhead{32 min} &
\colhead{45 min} &
\colhead{60 min} }
\startdata
1 & 0.745 & 0.315 & 0.188 & 0.120 \\
4 & 0.358 & 0.138 & 0.099 & 0.052 \\
8 & 0.253 & 0.087 & 0.051 & 0.032 \\
[-2pt]
\enddata
\tablecomments{See Section~\ref{sec:sens} for a discussion of this comparison of relative sensitivity to a highly accelerated fast-spinning binary pulsar.}
\end{deluxetable}

As expected, the power recovered in successively higher harmonics decreases for each of the four surveys. The value of $\gamma^2_{2m}$ in the first harmonic is a reasonable proxy for binary detectability; that is, the higher the fraction of power that is recovered in an acceleration search, the more likely one is to detect the MSP in a given observation. For \psrf{}, with a large and rapidly changing acceleration, 15-minute observations are significantly better at recovering power from a range of harmonics than longer integrations. Comparing $\gamma_{2m}^2$ in the first harmonic between the 15-minute Arecibo observation and the next-longest (32-minute GBT) observation shows that Arecibo yields a signal that is more than twice the strength of the GBT's (not considering differences in telescope gain and system temperature). The difference becomes even more dramatic for successively longer observations. While longer integration times improve sensitivity, the effect is only proportional to the square root of the observation length, while telescope gain is a directly proportional parameter. One strategy to combat the effects of acceleration (useful for relatively bright MSPs) is to take a long observation and apply acceleration searches to small subsections of the data, as well as searching the entire observation.

The characteristics of the Arecibo telescope give it a two-fold advantage over similar instruments. First, it has a significantly better raw sensitivity than both the GBT and Parkes for similar \fermi{}-LAT source searches. We are therefore able to detect \emph{fainter} systems, even in the absence of considerations relating to binary systems. Second, its large gain allows for short observations, which in turn increases its sensitivity to \emph{highly accelerated} binaries, of which there are many among the \fermi{} source population. Short observations also allow us to split observing time over multiple epochs, rather than integrating for a long time at a single epoch, further increasing our ability to combat eclipses and scintillation. Though its declination range is limited, the Arecibo telescope's raw sensitivity firmly establishes its indispensability as an MSP-finding resource.

\acknowledgements
H.\ Thankful Cromartie would like to thank the NSF, Fernando Camilo, the staff of the Arecibo Observatory, and its resident scientists for the opportunity to pursue this research, and for an experience that compelled her to continue in astrophysics.

The Arecibo Observatory is operated by SRI International under a cooperative agreement with the National Science Foundation (AST-1100968), and in alliance with Ana G. M\'endez-Universidad Metropolitana, and the Universities Space Research Association.

The National Radio Astronomy Observatory is a facility of the National Science Foundation operated under cooperative agreement by Associated Universities, Inc.

The \textit{Fermi} LAT Collaboration acknowledges generous ongoing support from a number of agencies and institutes that have supported both the development and the operation of the LAT as well as scientific data analysis. These include the National Aeronautics and Space Administration and the Department of Energy in the United States, the Commissariat \`a l'Energie Atomique and the Centre National de la Recherche Scientifique / Institut National de Physique Nucl\'eaire et de Physique des Particules in France, the Agenzia Spaziale Italiana and the Istituto Nazionale di Fisica Nucleare in Italy, the Ministry of Education, Culture, Sports, Science and Technology (MEXT), High Energy Accelerator Research Organization (KEK) and Japan Aerospace Exploration Agency (JAXA) in Japan, and the K.~A.~Wallenberg Foundation, the Swedish Research Council and the Swedish National Space Board in Sweden. Additional support for science analysis during the operations phase is gratefully acknowledged from the Istituto Nazionale di Astrofisica in Italy and the Centre National d'\'Etudes Spatiales in France.

{\em Facilities:} \facility{{\em Fermi} (LAT)}, \facility{Arecibo
Observatory (PUPPI)}, \facility{GBT (GUPPI)}

\end{document}